# Laser-Driven Growth of Semiconductor Nanowires from Colloidal Nanocrystals via the Young-Laplace Effect

**Authors:** Elena P. Pandres[a,1], Matthew J. Crane[a,1], E. James Davis[a], Peter J. Pauzauskie[a,b,c,d,*], and Vincent C. Holmberg[a,c,e,*]

**Affiliations:**

[a]Department of Chemical Engineering, University of Washington, Seattle, WA 98195-1750, United States

[b]Department of Materials Science and Engineering, University of Washington, Seattle, WA 98195-2120, United States

[c]Molecular Engineering & Sciences Institute, University of Washington, Seattle, WA 98195-1652, United States

[d]Physical and Computational Sciences Directorate, Pacific Northwest National Laboratory, Richland, WA 99352, United States

[e]Clean Energy Institute, University of Washington, Seattle, WA 98195-1653, United States

[1]These authors contributed equally to this work

* Correspondence to holmvc@uw.edu and peterpz@uw.edu

**Abstract**

The ability to produce nanowires through vapor- and solution-based processes has propelled nanowire material systems toward a wide range of technological applications. Conventional, vapor-based nanowire syntheses have enabled precise control over nanowire composition and phase. However, vapor-based nanowire growth employs batch processes with specialized pressure management systems designed to operate at high temperatures, limiting throughput. More recently developed solution-based nanowire growth processes have improved scalability but can require even more extensive pressure and temperature management systems. Here, we demonstrate a continuous-flow, solution-based nanowire growth process that utilizes the large Young-Laplace interfacial surface pressures and collective heating effects of colloidal metal nanocrystals under irradiation to drive semiconductor nanowire growth photothermally without the need for high-pressure or high-temperature equipment. In this process, a laser irradiates a solution containing metal nanocrystals and semiconductor precursors. Upon light absorption, the metal nanocrystals heat rapidly, inducing semiconductor precursor decomposition and nanowire growth. This process is performed on a benchtop in simple glassware under standard conditions. To demonstrate the generality of this technique, we synthesized three distinct semiconductor nanowire material systems: bismuth-seeded germanium nanowires, bismuth-seeded cadmium selenide nanowires, and indium-seeded germanium nanowires. The simplicity and versatility of this process opens the door to a range of experiments and technologies including in-line combinatorial identification of optimized reaction parameters, *in situ* spectroscopic measurements to study solution-based nanowire growth, and the potential production of nanowires with complex compositions or rationally incorporated dopants.

**Significance Statement**

Fundamental studies of semiconductor nanowire syntheses have enabled the production of nanowires with unique properties for applications in energy conversion, optoelectronics, and energy storage. However, these studies have focused on complex syntheses that are challenging to scale. Despite the potential scalability of solution-phase growth, the fundamental studies that are necessary to engineer growth and target desirable properties remain challenging. Here, we highlight a new method using the laser irradiation of colloidal nanocrystals to generate heat and synthesize semiconductor nanowires in optically interrogable reactors on the benchtop without complicated temperature or pressure management systems. This enables a new experimental platform for fundamental studies, rapid reactions for combinatorial screening, and the potential to realize atomically precise nanowire compositions.

**Introduction**

The absorption of light in tandem with subsequent energy transfer processes provides a unique handle to initiate chemical reactions in a wide range of materials, enabling key processes with diverse applications. For example, light absorption by photosensitive molecules has been used to engineer advanced biomaterials (1–5) as well as to initiate surface reactions, which impart additional functionality to inorganic materials(6, 7). In each of the above methods, the energy from the incident radiation is transferred through a photoinitiator to start the chemical reaction or a certain wavelength is chosen to target a specific bond. However, light-driven chemical and physical processes are not limited to photosensitizing molecules for energy transfer or to direct bond cleavage; inorganic nanomaterials also participate in light-driven energy transfer processes that can be used to drive chemical reactions(8–17) and phase transformations(18–24). For instance, the solar irradiation of plasmonic colloidal metal nanoparticles dispersed in water has been used to generate steam for off-grid water purification and medical instrument sterilization(18–20, 25–27). As another example of the extreme local conditions that can be created under such energy transfer processes, gold nanoparticles can be reshaped under resonant irradiation and have been shown to reach temperatures well above the boiling point of water(28, 29). In fact, photothermal heating of dilute nanoscale materials can locally superheat water to temperatures greater than 300°C without boiling due to the large Young-Laplace interfacial pressures that suppress bubble nucleation on highly curved nanoscale surfaces (30–32). This important thermodynamic effect – where the Young-Laplace interfacial pressure, $\Delta P$, is defined by the interfacial energy, $\gamma$, times the harmonic sum of the principle axes of curvature $(1/r_1 + 1/r_2)$ of the interface (*e.g.* $\Delta P = 2\gamma/r$ for a sphere) – can lead to a much wider range of accessible temperatures to drive liquid-phase chemistry at nanoscale interfaces. While conventional reactor systems are typically heated via conduction across the interface between the reactor sidewall and the fluid, when the heat source is decoupled from the reactor sidewalls, bubble nucleation can be avoided, thereby increasing the accessible temperature range at standard pressures. Taking advantage of the nanoscale Young-Laplace effect, the irradiation of metal nanocrystals has been used to induce high-temperature reaction conditions at nanocrystal surfaces in a low-cost manner, fabricate crystalline inorganic nanomaterials, and weld heterojunctions between nanowires (24, 33–35).

Importantly, metal nanocrystals are also essential for the growth of anisotropic semiconductor nanowires(36–41), which have great potential in optoelectronic(42, 43), sensing(44, 45), energy conversion(46–48), and energy storage applications(49–51). Typically, semiconductor nanowires are grown via decomposition of a semiconductor precursor and subsequent diffusion of semiconductor atoms into a metal nanocrystal seed, which, upon saturation, facilitates nucleation and continued growth of an anisotropic, crystalline semiconductor nanowire(52–54). Although semiconductor nanowire syntheses typically involve the global heating of a reactor, some reports have used metal nanocrystals not only as seeds for semiconductor nanowire growth, but also as local heat sources under resonant irradiation for precursor decomposition and diffusion(15–17). For example, irradiating metal nanoparticle-decorated substrates in a modified chemical vapor deposition chamber enabled site-specific heating and, ultimately, plasmon-assisted growth of nanowire arrays(15) as well as single nanowires(16, 17).

To date, photothermally-driven nanowire growth has largely been studied in substrate-based systems, yet the growth of semiconductor nanowires in dispersion through direct laser irradiation of colloidal metal nanocrystals remains largely unexplored.

A handful of reports have utilized substrate-bound metal nanocrystals to transduce incident irradiation into thermal energy to decompose semiconductor precursors for nanowire growth using a solution-based carrier phase(55–58); however, for both vapor- and solution-based nanowire growth, past work has focused primarily on irradiating a substrate decorated with metal nanocrystals(15–17, 55–57). While this substrate-based configuration exploits the photothermal heating of metal nanocrystals, it lacks the potentially advantageous scalability associated with solution-phase nanowire growth, including superheating via the Young-Laplace effect. Actualizing light-driven, substrate-free, solution-based nanowire growth would enable continuous flow-based nanowire growth on the benchtop as well as broader access to nanowire growth chemistries that require high temperatures without the need for expensive or niche equipment. Here, we capitalize on the unique properties of metallic colloidal nanoparticles under laser irradiation to act as sources of photothermal energy transduction as well as growth-directing seeds for the synthesis of anisotropic semiconductor nanostructures through a one-step, bottom-up, colloidal, solution-based process. By using colloidal metal nanocrystals to convert incident photons into thermal energy, as opposed to relying on global heating of the entire reactor system, we relax reactor design constraints to enable the continuous growth of semiconductor nanowires on the benchtop, eliminating the need for specialized, high-temperature or high-pressure equipment. By using a dispersion of nanocrystals, rather than a metal nanocrystal-decorated substrate, we demonstrate a system in which colloidal nanocrystals irradiated with a collimated infrared laser can reach the high temperatures (>200°C) required for solution-phase nanowire synthesis due to a combination of the nanoscale Young-Laplace effect and the collective heating effects of colloidal metallic nanoparticles. This demonstrates a generalizable process for semiconductor nanowire growth that can be performed on a benchtop in either batch or continuous operation, for potential rapid, high-throughput screening and parameter optimization of semiconductor nanowire growth.

**Results and Discussion**

This contact-free, laser-driven, solution-based nanowire growth process is detailed in Fig. 1, depicting a cuvette containing a dispersion of metal nanocrystals and molecular precursors (Fig. 1i) in an organic solvent. Upon irradiation (Fig. 1ii), the nanocrystals transduce the incident photons into thermal energy, which drives the decomposition of molecular semiconductor precursors, thereby inducing the formation of a saturated metal-semiconductor alloy that enables seeded nanowire growth (Fig. 1iii).

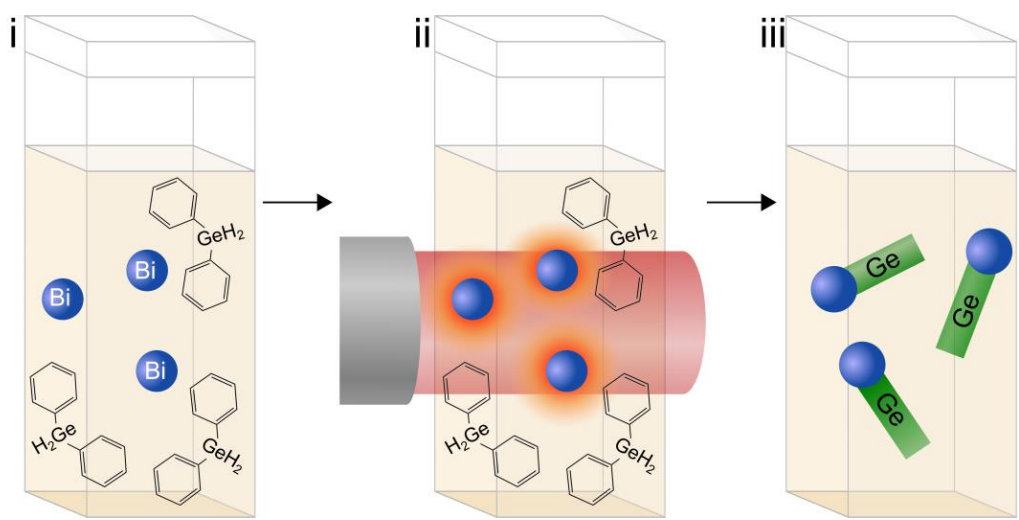

**Figure 1.** Illustration of contact-free, laser-driven, colloidal semiconductor nanowire growth in a high boiling point organic solvent on the benchtop in a quartz cuvette. (i-iii) The irradiation of colloidal metal nanocrystals with a collimated laser enables accumulation of thermal energy in solution, which can then be used to drive chemistry – such as the decomposition of molecular precursors – thus facilitating the contact-free, solution-based, photothermal synthesis of semiconductor nanowires.

While numerous studies have examined the heating of metal nanocrystals under irradiation, most investigations have focused on noble metals, such as gold(18–22, 30, 31). To examine the potential heating effects enabled by contact-free irradiation of the low-melting point colloidal metal nanocrystals used in this study (SI Appendix Fig. S1), we irradiated bismuth nanocrystals at a range of concentrations (Fig. 2, SI Appendix Fig. S2 a-d), denoted by the optical density of the dispersion at the 1070-nm excitation wavelength. Thermographic infrared images of a bismuth nanocrystal dispersion in octadecene ($OD_{1070}$ = 0.52) in a cuvette under irradiation (Fig. 2a-d, SI Appendix Movie S1) demonstrate the photothermal transduction characteristics of the colloidal nanocrystals, as compared to neat solvent under identical excitation conditions (Fig. 2e-h, SI Appendix Movie S1). As the optical density of the Bi nanocrystal dispersion increases, the maximum temperature achieved also increases by up to 170°C (Fig. 2i, Fig. S2 a-d), while the neat solvent temperature increases by a maximum of only 17°C over the same time period. Therefore, the observed temperature increase of the nanocrystal dispersions is due to photothermal transduction by the bismuth nanocrystals. We note that the thermographic images only reflect the temperature of the cuvette surface because the walls of the quartz cuvette absorb the long-wavelength infrared light used in thermographic imaging. Thus, the reaction volume is likely at a higher temperature due to the heat transfer resistance of both the solvent volume and the quartz cuvette walls.

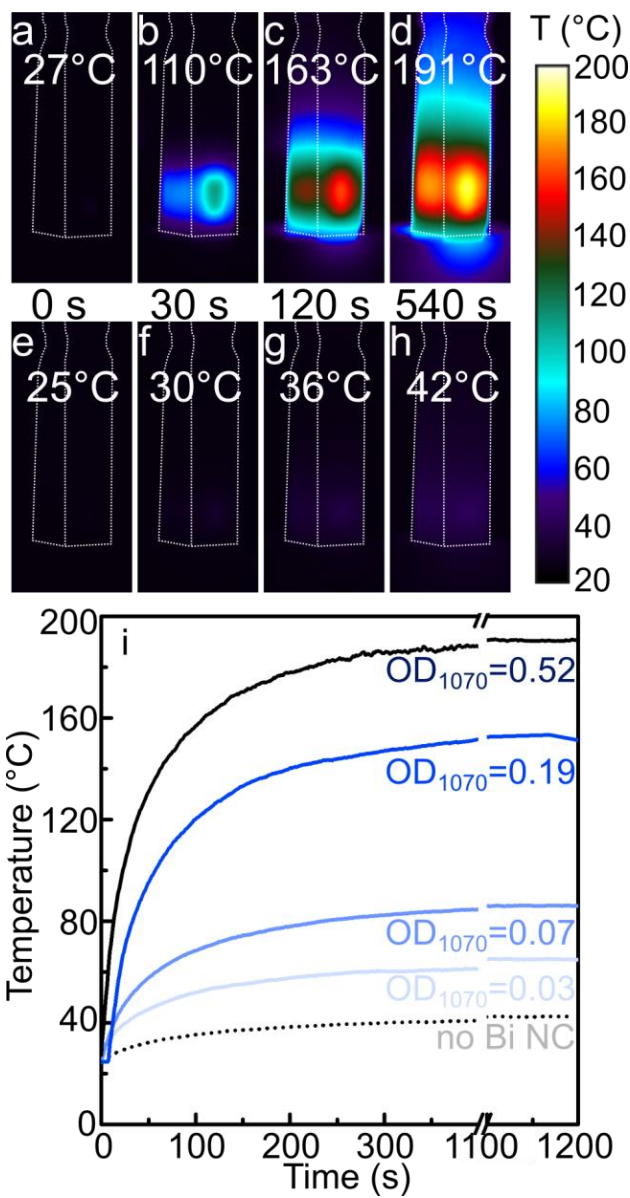

**Figure 2.** (a-d) Thermographic infrared images illustrating the evolution of temperature during irradiation of a bismuth nanocrystal dispersion ($OD_{1070}$ = 0.52) in a quartz cuvette with a near-infrared laser (1070 nm, 15 W), along with (e-h) a corresponding set of control images recorded for neat octadecene under identical irradiation conditions (1070 nm, 15 W) in the absence of nanocrystals (Movie S1). Temperature labels correspond to the maximum temperature recorded in each image. (i) Time-dependent temperature profiles for a series of bismuth nanocrystal concentrations ($OD_{1070}$ = 0.03-0.52) under 1070-nm irradiation (15 W).

In addition to measuring the time dependence of the temperature profile on the bismuth nanocrystal concentration, we used COMSOL Multiphysics software to simulate the surface temperature of an irradiated cuvette containing a bismuth nanocrystal dispersion (SI Appendix Fig. S3 a-d). These numerical solutions accurately predict the surface temperature of the cuvette at low bismuth nanocrystal concentrations; however, the predicted temperatures diverge from the experimentally measured values at higher nanocrystal concentrations and higher temperatures. The divergent numerical solutions at higher bismuth nanocrystal concentrations are likely due to high-temperature effects, such as solvent boiling and reflux, Ostwald ripening, or coalescence of bismuth nanocrystals, which are not included in the model. With these caveats, the numerical model provides an upper bound for the internal temperature profiles and demonstrates that the nanocrystals in the center of the cuvette reach high temperatures under laser irradiation (SI Appendix Fig. S3 e-f). Using an analytical solution based on a Mie theory source term(59), we also calculated the temperatures of a single bismuth or indium nanocrystal (discussed later) suspended in an infinite bath of the growth solution. In this formalism, isolated bismuth and indium nanocrystals heat very little relative to the overall growth solution. Computations predict that at these experimental laser irradiances, a single bismuth or indium nanocrystal cannot heat sufficiently to decompose molecular precursors and drive nanowire growth (SI Appendix Fig. S9). Critically, this observation implies that the nanocrystals heat *via* the collective scattering and absorption of photons. This is in agreement with previous work performed with gold nanoparticles in an aqueous solution under irradiation(12, 19, 25). Thus, reaching the threshold temperature for nanowire growth depends on many factors, including the nanocrystal concentration, the thermal properties of the dispersion, and the irradiance of the incident laser.

We next incorporated II-VI molecular precursors into the colloidal nanocrystal solution in an attempt to drive solution-liquid-solid (SLS) nanowire growth *via* photothermal heating. In a typical laser-driven synthesis, a solution of colloidal nanocrystals and molecular precursors in a cuvette were irradiated with a collimated laser. The optical density of the dispersion was observed to increase dramatically within the first 30 seconds of irradiation. When the laser was blocked, the reaction product rapidly flocculated, indicating a substantial change in morphology relative to the starting nanocrystals. Bright-field transmission electron microscopy (TEM) imaging (Fig. 3b-d) clearly demonstrates that laser heating did not result in homogeneously nucleated CdSe byproducts upon precursor decomposition. Rather, the bismuth nanocrystal seeds act as growth-directing agents to produce CdSe nanowires via a photothermally driven SLS growth mechanism. Compared to previous two-step, optically driven II-IV nanowire growth processes(57), the CdSe nanowire synthesis performed here was carried out in a single step. TEM images (Fig. 3b-d) highlight the narrow nanowire diameters (7.3 ± 1.9 nm, Fig. S5) achieved via this photothermally driven process. High-resolution TEM imaging (Fig. 3d) shows that the CdSe nanowires exhibit the <0001> growth direction that is typical of anisotropic wurtzite-phase CdSe. X-ray diffraction (XRD) indicates that the CdSe nanowires are a mixture of wurtzite and zinc-blende phases, as is commonly observed in most CdSe nanowire syntheses; the mixture of the two phases can be attributed to the minimal energy difference between <111> zinc-blende and <0001> wurtzite nanowire growth (60). Notably, XRD also shows an increased intensity of the (002) and (111) reflections for wurtzite and zinc-blende, respectively, which is consistent with previously reported growth directions associated with each phase of the anisotropic nanowire (Fig. 3e) (60). Control experiments based on the irradiation of CdSe molecular precursor solutions in the absence of bismuth nanocrystals resulted in minor temperature increases (SI Appendix Fig. S2e) and no detectible precursor decomposition or nanowire growth. These results clearly demonstrate that under irradiation, the bismuth nanocrystal dispersion heats rapidly, facilitating decomposition of the cadmium and selenium molecular precursors and subsequent solution-based growth of bismuth-seeded CdSe nanowires.

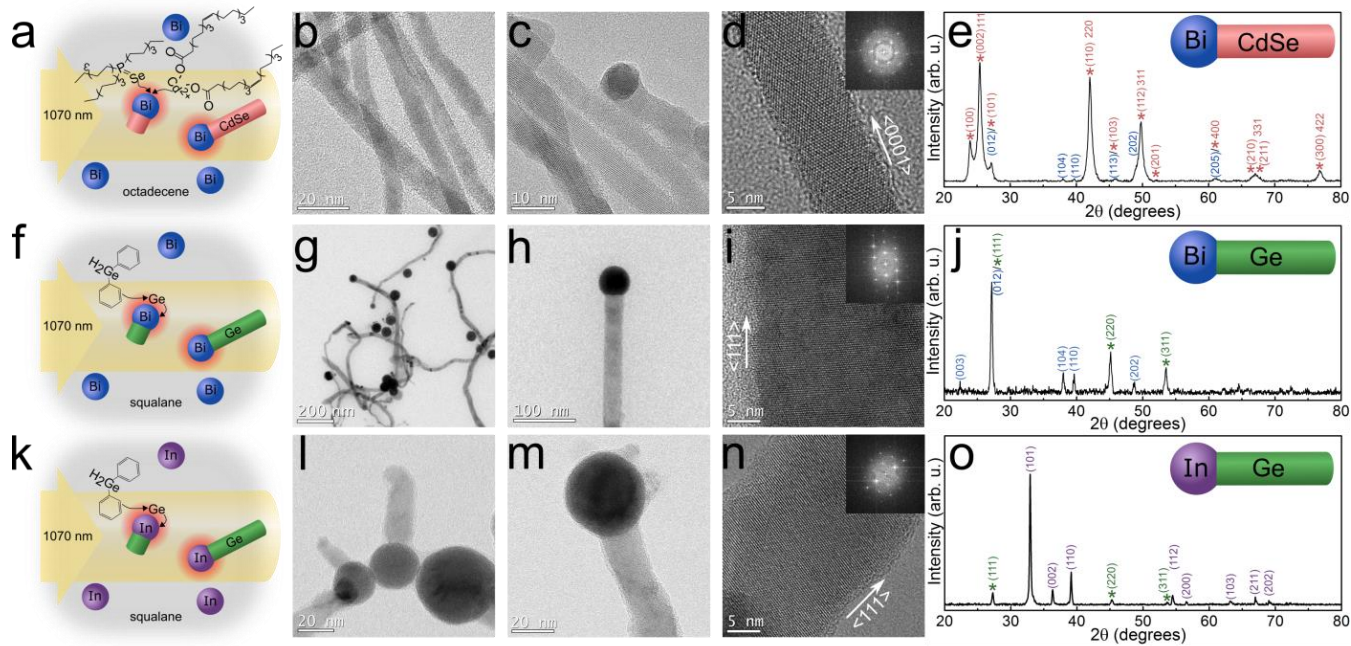

**Figure 3.** Examples of contact-free, solution-based, laser-driven nanowire growth on the benchtop for nanowire systems with both ionic and covalent character, produced using different metal nanocrystal seeds under 1070-nm excitation. (a-e) Narrow-diameter, polytypic wurtzite/zinc-blende CdSe nanowires (hexagonal reflection indices denoted in parentheses) grown from Bi nanocrystal seeds using cadmium oleate and trioctylphosphine selenide precursors. (f-j) Diamond-cubic, Bi-seeded Ge nanowires produced via photothermal decomposition of diphenylgermane. (k-o) In-seeded Ge nanowires grown using In nanocrystal seeds as photothermal heat sources. Both nanowire (highlighted with asterisks) and nanocrystal seed materials are indexed in each X-ray diffraction pattern.

Notably, typical nanocrystal-seeded, SLS-based CdSe nanowire growth processes require the use of a Schlenk line and reaction temperatures in the range of 230-350°C. Here, heat transduced by the nanocrystals in the contact-free, photothermal growth reaction enabled the same high temperature process to be carried out in a standard quartz cuvette on a benchtop. This is emphasized by the observation that the $OD_{1070} = 0.07$ Bi nanocrystal dispersion resulted in the successful growth of CdSe nanowires, despite the fact that the surface of the cuvette only reached a maximum temperature of 136°C (SI Appendix Fig. S2g). This result further indicates that the temperature at the nanocrystal-solvent interface is higher than the measured surface temperature of the cuvette. In addition to using a quartz cuvette as a reaction vessel, we have also successfully employed a simple NMR tube combined with a low-cost, low-power (2 W), 808-nm diode to grow Bi-seeded CdSe nanowires (Movie S2), demonstrating that optically driven nanowire growth can be carried out using a variety of excitation powers and wavelengths.

While this represents a successful demonstration of solution-based, photothermally driven seeded nanowire growth, CdSe has a large degree of ionic character, and material systems that are more covalent (such as Si and Ge) typically have much higher crystallization barriers, and, therefore, require significantly higher temperatures (>300°C) for nanowire growth(61–66). Consequently, in order to investigate the versatility and potential scope of the photothermally driven nanowire growth process, we also targeted the production of group-IV semiconductor nanowires, using the same bismuth nanocrystal seeds to drive the reaction. In this scenario, colloidal bismuth nanocrystals under near-infrared irradiation once again act both as heat sources and as seeds for nanowire growth, driving the decomposition of diphenylgermane precursor to facilitate the optically driven SLS-growth of Ge nanowires (Fig. 3f-j) without the need for a vacuum system, Schlenk line, or supercritical reactor system. Bright-field TEM images (Fig. 3g-

i) demonstrate the growth of crystalline Ge nanowires with diameters of 15.6 ± 7.8 nm (SI Appendix Fig. S6) and a <111> growth direction, which is characteristic of high-temperature growth of Ge nanowires (67). While XRD (Fig. 3f) clearly shows the presence of both crystalline Bi and Ge, the Bi-seeded Ge nanowires synthesized through contact-free photothermal heating were typically observed to have a tortuous morphology, likely due to the small volume of solution that was irradiated, and subsequent convection of nanowires into and out of the beam.

In order to demonstrate the generality of this approach, we next investigated whether indium nanocrystals could be used for laser-driven nanowire growth. While indium has been used to grow germanium nanowires via vapor-liquid-solid growth(68) and other types of semiconductor nanowires through modified SLS growth(63), previous work has primarily employed the use of evaporated In thin films to seed growth. Here, we demonstrate the use of colloidal In nanocrystals (SI Appendix Fig. S7) as photothermal heat transducers under incident near-infrared irradiation to decompose diphenylgermane in solution, once again enabling the laser-driven, seeded growth of germanium nanowires in a cuvette (Fig. 3k-o). The In-seeded Ge nanowires generated from this optically driven process typically had shorter lengths and wider diameters (Fig. 3k-o, Fig. S8) than the photothermally grown Bi-seeded Ge nanowires. These morphological differences can likely be attributed to the markedly lower melting point of In (156°C) relative to Bi (271°C), which could result in rapid coalescence of In nanoparticles upon heating, leading to larger seed particles and, thus, larger nanowire diameters. The differences in surface energies and contact angles for the two different nanocrystal seeds could also contribute to differences in morphology.

All of these processes generated SLS-grown nanowires at ambient temperatures via photothermal heating without the use of a Schlenk line, pressure vessel, insulation, or resistive heating. Specifically, the high temperatures generated by contact-free, optical heating of the metal nanocrystals rapidly drive nanowire growth without the use of any high-temperature equipment. In comparison, global resistive heating during a conventional SLS nanowire growth reaction involves an isothermal process in which the reaction vessel and chemical components are at thermal equilibrium.

The versatility of this growth strategy naturally lends itself toward the integration of spectroscopic methods that could be used to study *in situ* nanowire growth dynamics and to rapidly tune reaction conditions. However, in all of the above nanowire growth demonstrations—using both quartz cuvettes and NMR tubes as reaction vessels—the nanowire growth solution inevitably developed convection currents due to the volume of the photothermal heating zone and variations in intensity across the Gaussian laser source relative to the solution volume. To realize the potential of this contact-free, colloidal nanocrystal-based semiconductor nanowire growth process, as a proof of principle, we demonstrated the photothermally driven growth of semiconductor nanowires using a continuous-flow reactor to circumvent problems associated with free convection and to validate laser-driven nanowire growth for future rapid screening of synthesis parameters (Fig. 4). We first constructed an optically accessible reaction zone (Fig. 4a), through which solutions of II-VI semiconductor precursors and colloidal nanocrystals can be injected simultaneously. Upon irradiation and while under flow, the nanocrystals once again rapidly generated heat to decompose the precursors and seed nanowire growth (Fig. 4d-e), eventually exiting the flow cell further downstream. Converting this non-contact, laser-driven nanowire growth method into a continuous flow process is a further step toward an easily-accessible, low-cost, scalable system in which nanowire growth dynamics and various growth parameters can be rapidly scanned and spectroscopically probed to enable real-time feedback and optimization of material properties.

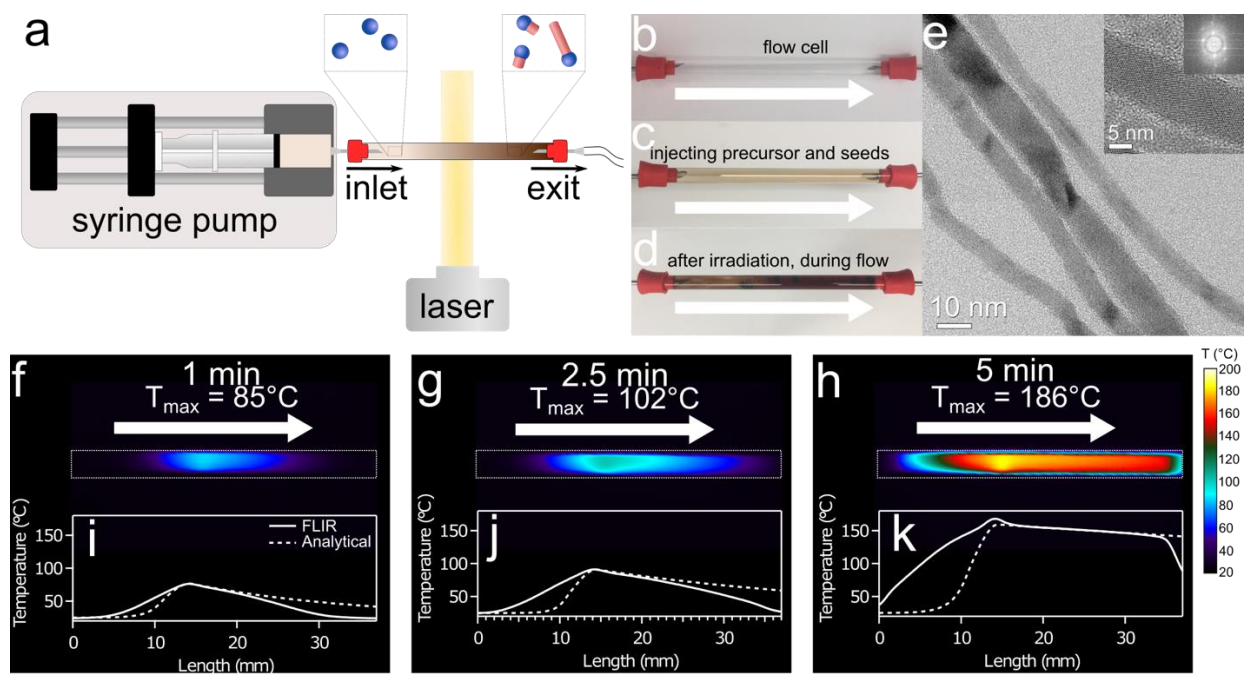

**Figure 4.** Demonstration of continuous, laser-driven nanowire growth under flow. (a) Flow cell schematic along with (b-d) associated images of the reaction zone (b) containing only octadecene, (c) during injection of Bi nanocrystal seeds ($OD_{1070}$ = 0.07) and II-VI molecular precursors, and (d) after irradiation under flow. (e) Characteristic TEM images of the Bi-seeded CdSe nanowire product. (f-h) Snapshots from an infrared thermal imaging camera, showing spatial and temporal evolution of temperature profiles of the flow cell after 1 minute, 2.5 minutes, and 5 minutes of 1070-nm irradiation under continuous flow (0.05 mL/min). Temperature labels correspond to the maximum temperature recorded in each image. (i-k) A comparison of the radially averaged FLIR temperatures (solid line) and analytically predicted temperatures (dashed line) of the flow cell after 1, 2.5, and 5 minutes of photothermal heating.

To determine the internal temperature of the flow cell during heating, we analytically calculated the temperature profile within the reactor volume during laser-driven nanowire growth under flow(69). We include details of the derivation and model, which should be valid for a wide range of flow rates, in the SI Appendix (70). Briefly, we start from the energy equation,

$$2\rho_s C_s v_z \frac{\partial T}{\partial z} = k_s \left[\frac{1}{r}\frac{\partial}{\partial r}\left(r \frac{\partial T}{\partial r}\right) + \frac{\partial^2 T}{\partial r^2}\right] + S, \tag{1}$$

where $\rho_s$, $C_s$, and $v_z$ are the density, heat capacity, and velocity of the solution in the flow reactor, $T$, $z$, and $r$ are the temperature and axial and radial coordinates, and $S$ is the source term. We solve the equation for three different regions: upstream of the irradiated region (S = 0), the region under laser irradiation (S > 0), and downstream of the laser irradiation (S = 0). At axial interfaces between these three regions, we match the temperatures and flux; at axial interfaces with inflow and outflow, we ensure the temperature is bounded as $z \to \pm\infty$; for radial interfaces, we use a Robin's boundary condition based on correlations for natural conduction from a horizontal cylinder(71).

At low flow rates, that is for small Peclet numbers (Pe < 10), heat transport via both thermal conduction and thermal convection are important and cannot be neglected. The Peclet number is defined by $Pe = R_1 \langle v_z \rangle / \alpha_s$ in which $R_1$ is the inner radius of the tube, and $\alpha_s$ is the thermal diffusivity of the fluid ($\alpha_s = k_s/\rho_s C_s$). Thermal camera images (Fig. 4f-h) of the flow reactor clearly demonstrate axial diffusion of heat both upstream and downstream of the incident laser spot experimentally verifying that axial conduction is taking place. Figure 4 (panels i-k) shows the projected temperature distribution on the surface of the flow cell and the analytically calculated, radially averaged temperature. The analysis reveals that there is a ~1% gradient between the center of the reactor fluid volume and the surface of the chamber. This lack of a radial thermal gradient suggests that the maximum temperature during growth in the flow cell was ~188°C based on thermal camera images, which is less than the typical 230°C temperature used in bismuth-seeded CdSe nanowire synthesis(72). We note that analytical temperature distributions of both a bismuth-seeded CdSe nanowire and a bismuth nanocrystal (SI Appendix Fig. S9 and S10) in an infinite bath do not demonstrate appreciable heating above ambient temperatures(24). It is possible that hot-electron effects or exothermic precursor decomposition may contribute to the energy required for semiconductor nanowire growth. To examine these possibilities, we photothermally heated colloidal bismuth nanocrystals in a cuvette and measured the surface temperatures with and without CdSe precursors (SI Appendix Fig. S2d and Fig. S2h). After nine minutes, the surfaces of the cuvettes containing the irradiated nanocrystal dispersions with CdSe precursors present were 96°C higher in temperature than cuvettes without semiconductor precursors. This additional heating may be associated with an exothermic reaction of CdSe precursors, which may accelerate nanowire synthesis; alternatively, the growth of nanowires may generate additional photon scattering or absorption effects that could boost the absorption of light, thus generating more heat. The analytically calculated temperature distributions for the flow reactor best match the measured temperature when the absorption efficiency and solution thermal conductivity increase over time, which would occur if either the growth of nanowires increases the absorption coefficient of the solution(73, 74)—simultaneously increasing the thermal conductivity—and/or if the reaction is exothermic, supporting the above hypotheses (SI Appendix Table S1).

**Conclusions**

We have demonstrated that photothermal heating of colloidal metal nanocrystals via laser irradiation is a versatile approach to drive the solution-based synthesis of semiconductor nanowires with both ionic and covalent character, demonstrating the versatility of this method. In these syntheses, Bi and In metal nanocrystals were used as contact-free photothermal transducers that facilitate the rapid SLS growth of semiconductor nanowires. Because these nanocrystals are dispersed in solution, rather than pinned to a substrate, they have the potential to locally superheat the solvent due to the large Young-Laplace interfacial surface pressure required to nucleate a bubble. Consequently, both temperature and gas handling constraints of the reactor are relaxed, enabling the use of low-cost quartz cuvettes or NMR tubes as benchtop reactors for nanowire synthesis. These results enable the translation from a batch process to a scalable, continuous-flow process. With further optimization, this optically driven synthesis could enable the solution-based growth of nanowires that currently cannot be synthesized in solution due to solvent boiling point limitations. In addition, photothermal heating makes rapid changes in temperature possible, which could facilitate the production of complex, heterostructured nanowires or rational dopant incorporation via solution-based methods. The ability to quickly reach high temperatures near the metal nanocrystal seed could potentially enable the nonequilibrium incorporation of atomically precise semiconductor dopants during the nanowire growth process, rather than during an additional post-synthetic doping step. Moreover, because this process can be performed on a benchtop in virtually any chemically inert, optically accessible reaction vessel, it could enable the direct integration of *in situ* scattering and spectroscopic techniques to study solution-based nanowire growth dynamics, rapid combinatorial scanning of new nanowire chemistries, and optimize nanowire growth conditions with real-time feedback. This laser-driven nanowire growth strategy may also enable the study of single-particle nanowire growth dynamics through scattering measurements or the combination of *in situ* optical spectroscopy with single-beam laser tweezers(24).

**Materials**

All chemicals were used as received unless otherwise noted. Bismuth (III) chloride ($BiCl_3$, 99.99% trace metals basis), cadmium oxide (CdO, ≥99.9% trace metals basis), ethanol (anhydrous,

≤0.005% water), indium (III) chloride (99.999% trace metals basis), isopropanol (anhydrous, 99.5%), molecular sieves (3 Å), *n*-butyllithium (1.3 M in *n*-heptane), 1-octadecene (ODE, 90%), oleylamine (90%), oleic acid (degassed, 90%), sodium bis(trimethylsilyl)amide (1.0 M solution in tetrahydrofuran), squalane (96%), Super-Hydride® solution (1.0 M lithium triethylborohydride in tetrahydrofuran), selenium (Se, <5 mm particle size, ≥99.999% trace metals basis), tetrahydrofuran (THF, anhydrous, ≥99.9%), toluene (anhydrous, 99.8%), and trioctylphosphine (TOP, 97%) were purchased from Sigma. Diphenylgermane (>95%) was purchased from Gelest. Poly(1-hexadecene-*co*-1-vinylpyrrolidinone) (PHD-*co*-PVP) was provided by Ashland under the trade name Ganex™ V-216. Toluene (Certified ACS, 99.8%) was purchased from Fisher. Ethanol (200 proof) was purchased from Decon Laboratories.

**Methods**

Bismuth Nanocrystal Synthesis

Bismuth nanocrystals were synthesized based on the protocol outlined by Wang, *et al.*[1] A 25 wt.% solution of PHD-*co*-PVP in 1-octadecene was dried over molecular sieves for one week. BiCl$_3$ ($2.6 \times 10^{-2}$ mmol) was mixed at 800 RPM with THF (570 μL) under nitrogen for thirty minutes. The solution of PHD-*co*-PVP in octadecene (6.1 mL) was added to the flask under continuous mixing at 900 RPM. The flask was cycled between nitrogen and vacuum three times. A solution of 1.0 M sodium bis(trimethylsilyl)amide in THF (825 μL) was injected, and the flask was stirred until a dark orange-brown color was observed (~10 minutes, 1100 RPM). The solution was then heated to 200°C for 17 hours, cooled, and then transferred into a nitrogen-filled glove box for storage. Bismuth nanocrystals were handled under nitrogen for cleaning. Nanocrystals were washed five times using a 1:4 ratio of anhydrous hexanes:ethanol and centrifugation (6800 RCF for 10 minutes) to remove excess PHD-*co*-PVP prior to use for nanowire growth. The average diameter of the bismuth nanocrystal seeds was 52 ± 8 nm.

Indium Nanocrystal Synthesis

Indium nanocrystals were synthesized using the protocol outlined by He, *et al*.[2] $InCl_3$ ($1.4\times10^{-1}$ mmol) was measured in the glove box, transferred into a three-neck flask, and attached to a Schlenk line. The flask was purged with nitrogen and then transitioned between vacuum and nitrogen atmosphere three times. Oleylamine (13 mL) was added to the flask, which was heated to 100°C while mixing (1000 RPM) and then placed under vacuum for an additional 45 minutes. The mixture was blanketed with nitrogen and further heated to 160°C. A solution of 1.3 M *n*-butyllithium in *n*-heptane (1.3 mL) was injected into the flask, followed by an injection of 1.0 M Super-Hydride® solution (300 µL). The reaction was allowed to run for ten seconds, whereupon 12 mL of anhydrous toluene was injected to cool the solution and quench the reaction. At 50°C, oleic acid (400 µL) was added to stabilize the nanocrystals. Nanocrystals were then transferred into a glove box and handled under nitrogen for cleaning. Anhydrous ethanol (12 mL) was added to the dispersion, which was centrifuged at 4180 RCF for 10 minutes. The supernatant was discarded and the nanocrystals were washed with a 2:1 ratio of anhydrous hexanes:ethanol three more times. The average diameter of the indium nanocrystal seeds was 16 ± 2 nm.

Laser-Driven Nanowire Growth

Nanowire precursor growth solutions consisting of metal nanocrystals, molecular semiconductor precursors, and degassed solvents (squalane or ODE), were loaded into a screw-top quartz cuvette in the glove box, where the top of the cuvette was lined with Teflon and then wrapped with Parafilm to prevent oxidation of nanocrystals. Nanowire precursor growth solutions for each material system are described in the SI Appendix. The cuvette with the growth solution was transferred out of the glove box and subsequently irradiated with a polarized near-infrared fiber laser (λ=1070 nm) at a range of powers (15-70 W) and times (5-20 minutes). The resulting product was washed with a 2:1 ratio of toluene and ethanol, collected by centrifugation at 19,000 RCF, dispersed in toluene, and washed two more times. For details of photothermal nanowire growth under continuous flow, see the SI Appendix.

**Materials Characterization**

Transmission electron microscopy (TEM) images were acquired with a FEI Tecnai G2 F20 Supertwin TEM operating with a 200-kV accelerating voltage and were analyzed using ImageJ Software. X-ray diffraction (XRD) scans were collected using a Bruker D8 Discover equipped with an IμS 2-D XRD detector system and were analyzed using EVA software. Inductively coupled plasma optical emission spectroscopy (ICP-OES) was performed with a Perkin Elmer Optima 8300 spectrophotometer. UV-vis extinction spectra were collected using an Agilent Cary 60 UV-vis spectrophotometer. Time-dependent infrared thermal imaging was performed with a FLIR A325sc camera using the 0-350°C temperature range setting, and the resulting videos were analyzed using ResearchIR software.

**SI Appendix**

The SI Appendix contains additional information regarding experimental methods, analytical heat transport models of nanocrystals and nanowires, finite-element heat transport solutions, and the analytical heat transport derivation of the continuous flow reactor.

**Data Availability**

We have provided all relevant data in the text, figures, and SI Appendix.

**Author Contributions**

E.P.P. and M.J.C. performed the experiments and contributed equally to this work. M.J.C., P.J.P., and E.J.D. performed the heat transport analysis. V.C.H. and P.J.P. conceived the experiments and directed the research. All authors contributed to the design of experiments, scientific discussion, and writing of the manuscript.

**Acknowledgments**

This research was supported by the National Science Foundation (NSF) through the UW Molecular Engineering Materials Center, a Materials Research Science and Engineering Center (DMR-1719797), the University of Washington Molecular Engineering Institute, and the State of Washington through the University of Washington Clean Energy Institute and via funding from the Washington Research Foundation. Part of this work was conducted at the Molecular Analysis Facility, a National Nanotechnology Coordinated Infrastructure site at the University of Washington which is supported in part by the National Science Foundation (grant ECC-1542101), the University of Washington, the Molecular Engineering & Sciences Institute, and the Clean Energy Institute. M.J.C. acknowledges a NDSEG Fellowship as well as a WRF Postdoctoral Fellowship. P.J.P. acknowledges support from UW's Institute for Nano-engineered Systems (NanoES).

**References**
1. B. D. Fairbanks, M. P. Schwartz, C. N. Bowman, K. S. Anseth, Photoinitiated polymerization of PEG-diacrylate with lithium phenyl-2,4,6-trimethylbenzoylphosphinate: polymerization rate and cytocompatibility. *Biomaterials* **30**, 6702–6707 (2009).

2. D. L. Hern, J. A. Hubbell, Incorporation of adhesion peptides into nonadhesive hydrogels useful for tissue resurfacing. *J. Biomed. Mater. Res.* **39**, 266–276 (1998).

3. C. A. DeForest, K. S. Anseth, Cytocompatible click-based hydrogels with dynamically tunable properties through orthogonal photoconjugation and photocleavage reactions. *Nat. Chem.* **3**, 925–931 (2011).

4. A. M. Kloxin, A. M. Kasko, C. N. Salinas, K. S. Anseth, Photodegradable Hydrogels for Dynamic Tuning of Physical and Chemical Properties. *Science* **324**, 59–63 (2009).

5. B. A. Badeau, M. P. Comerford, C. K. Arakawa, J. A. Shadish, C. A. DeForest, Engineered modular biomaterial logic gates for environmentally triggered therapeutic delivery. *Nat. Chem.* **10**, 251–258 (2018).

6. J. M. Buriak, Illuminating Silicon Surface Hydrosilylation: An Unexpected Plurality of Mechanisms. *Chem. Mater.* **26**, 763–772 (2014).

7. J. A. Kelly, J. G. C. Veinot, An Investigation into Near-UV Hydrosilylation of Freestanding Silicon Nanocrystals. *ACS Nano* **4**, 4645–4656 (2010).

8. L. Zhou, *et al.*, Aluminum Nanocrystals as a Plasmonic Photocatalyst for Hydrogen Dissociation. *Nano Lett.* **16**, 1478–1484 (2016).

9. S. Mukherjee, *et al.*, Hot-Electron-Induced Dissociation of H2 on Gold Nanoparticles Supported on SiO2. *J. Am. Chem. Soc.* **136**, 64–67 (2014).

10. S. C. Jensen, S. Bettis Homan, E. A. Weiss, Photocatalytic Conversion of Nitrobenzene to Aniline through Sequential Proton-Coupled One-Electron Transfers from a Cadmium Sulfide Quantum Dot. *J. Am. Chem. Soc.* **138**, 1591–1600 (2016).

11. M. J. Enright, K. Gilbert-Bass, H. Sarsito, B. M. Cossairt, Photolytic C–O Bond Cleavage with Quantum Dots. *Chem. Mater.* **31**, 2677–2682 (2019).

12. H. Robatjazi, *et al.*, Plasmon-induced selective carbon dioxide conversion on earth-abundant aluminum-cuprous oxide antenna-reactor nanoparticles. *Nat. Commun.* **8**, 1–10 (2017).

13. D. Erickson, D. Sinton, D. Psaltis, Optofluidics for energy applications. *Nat. Photonics* **5**, 583–590 (2011).

14. J. R. Adleman, D. A. Boyd, D. G. Goodwin, D. Psaltis, Heterogenous Catalysis Mediated by Plasmon Heating. *Nano Lett.* **9**, 4417–4423 (2009).

15. D. A. Boyd, L. Greengard, M. Brongersma, M. Y. El-Naggar, D. G. Goodwin, Plasmon-Assisted Chemical Vapor Deposition. *Nano Lett.* **6**, 2592–2597 (2006).

16. L. Cao, D. N. Barsic, A. R. Guichard, M. L. Brongersma, Plasmon-Assisted Local Temperature Control to Pattern Individual Semiconductor Nanowires and Carbon Nanotubes. *Nano Lett.* **7**, 3523–3527 (2007).

17. G. Di Martino, F. B. Michaelis, A. R. Salmon, S. Hofmann, J. J. Baumberg, Controlling Nanowire Growth by Light. *Nano Lett.* **15**, 7452–7457 (2015).

18. O. Neumann, *et al.*, Solar Vapor Generation Enabled by Nanoparticles. *ACS Nano* **7**, 42–49 (2013).

19. Z. Fang, *et al.*, Evolution of Light-Induced Vapor Generation at a Liquid-Immersed Metallic Nanoparticle. *Nano Lett.* **13**, 1736–1742 (2013).

20. E. Lukianova-Hleb, *et al.*, Plasmonic Nanobubbles as Transient Vapor Nanobubbles Generated around Plasmonic Nanoparticles. *ACS Nano* **4**, 2109–2123 (2010).

21. S. Baral, A. J. Green, M. Y. Livshits, A. O. Govorov, H. H. Richardson, Comparison of Vapor Formation of Water at the Solid/Water Interface to Colloidal Solutions Using Optically Excited Gold Nanostructures. *ACS Nano* **8**, 1439–1448 (2014).

22. H. H. Richardson, *et al.*, Thermooptical Properties of Gold Nanoparticles Embedded in Ice: Characterization of Heat Generation and Melting. *Nano Lett.* **6**, 783–788 (2006).

23. R. Jin, *et al.*, Photoinduced Conversion of Silver Nanospheres to Nanoprisms. *Science* **294**, 1901–1903 (2001).

24. M. J. Crane, E. P. Pandres, E. J. Davis, V. C. Holmberg, P. J. Pauzauskie, Optically oriented attachment of nanoscale metal-semiconductor heterostructures in organic solvents via photonic nanosoldering. *Nat. Commun.* **10**, 1–7 (2019).

25. N. J. Hogan, *et al.*, Nanoparticles Heat through Light Localization. *Nano Lett.* **14**, 4640–4645 (2014).

26. O. Neumann, *et al.*, Compact solar autoclave based on steam generation using broadband light-harvesting nanoparticles. *Proc. Natl. Acad. Sci.* **110**, 11677–11681 (2013).

27. , Nanophotonics-enabled solar membrane distillation for off-grid water purification | PNAS (December 4, 2019).

28. W. Ni, H. Ba, A. A. Lutich, F. Jäckel, J. Feldmann, Enhancing Single-Nanoparticle Surface-Chemistry by Plasmonic Overheating in an Optical Trap. *Nano Lett.* **12**, 4647–4650 (2012).

29. A. Kuhlicke, S. Schietinger, C. Matyssek, K. Busch, O. Benson, In Situ Observation of Plasmon Tuning in a Single Gold Nanoparticle during Controlled Melting. *Nano Lett.* **13**, 2041–2046 (2013).

30. A. Kyrsting, P. M. Bendix, D. G. Stamou, L. B. Oddershede, Heat Profiling of Three-Dimensionally Optically Trapped Gold Nanoparticles using Vesicle Cargo Release. *Nano Lett.* **11**, 888–892 (2011).

31. M. T. Carlson, A. J. Green, H. H. Richardson, Superheating Water by CW Excitation of Gold Nanodots. *Nano Lett.* **12**, 1534–1537 (2012).

32. G. Baffou, J. Polleux, H. Rigneault, S. Monneret, Super-Heating and Micro-Bubble Generation around Plasmonic Nanoparticles under cw Illumination. *J. Phys. Chem. C* **118**, 4890–4898 (2014).

33. H. M. L. Robert, *et al.*, Light-Assisted Solvothermal Chemistry Using Plasmonic Nanoparticles. *ACS Omega* **1**, 2–8 (2016).

34. P. Christopher, H. Xin, S. Linic, Visible-light-enhanced catalytic oxidation reactions on plasmonic silver nanostructures. *Nat. Chem.* **3**, 467–472 (2011).

35. J. Qiu, W. D. Wei, Surface Plasmon-Mediated Photothermal Chemistry. *J. Phys. Chem. C* **118**, 20735–20749 (2014).

36. Y. Wu, P. Yang, Direct Observation of Vapor−Liquid−Solid Nanowire Growth. *J. Am. Chem. Soc.* **123**, 3165–3166 (2001).

37. A. M. Morales, C. M. Lieber, A Laser Ablation Method for the Synthesis of Crystalline Semiconductor Nanowires. *Science* **279**, 208–211 (1998).

38. W. Lu, C. M. Lieber, Semiconductor nanowires. *J. Phys. Appl. Phys.* **39**, R387–R406 (2006).

39. T. Måartensson, M. Borgström, W. Seifert, B. J. Ohlsson, L. Samuelson, Fabrication of individually seeded nanowire arrays by vapour–liquid–solid growth. *Nanotechnology* **14**, 1255–1258 (2003).

40. N. P. Dasgupta, *et al.*, 25th Anniversary Article: Semiconductor Nanowires – Synthesis, Characterization, and Applications. *Adv. Mater.* **26**, 2137–2184 (2014).

41. M. J. Crane, P. J. Pauzauskie, Mass Transport in Nanowire Synthesis: An Overview of Scalable Nanomanufacturing. *J. Mater. Sci. Technol.* **31**, 523–532 (2015).

42. L. N. Quan, J. Kang, C.-Z. Ning, P. Yang, Nanowires for Photonics. *Chem. Rev.* **119**, 9153–9169 (2019).

43. J. Huang, M. Lai, J. Lin, P. Yang, Rich Chemistry in Inorganic Halide Perovskite Nanostructures. *Adv. Mater.* **30**, 1802856 (2018).

44. Y. Cui, Q. Wei, H. Park, C. M. Lieber, Nanowire Nanosensors for Highly Sensitive and Selective Detection of Biological and Chemical Species. *Science* **293**, 1289–1292 (2001).

45. F. Patolsky, C. M. Lieber, Nanowire nanosensors. *Mater. Today* **8**, 20–28 (2005).

46. A. I. Hochbaum, P. Yang, Semiconductor Nanowires for Energy Conversion. *Chem. Rev.* **110**, 527–546 (2010).

47. I. Åberg, *et al.*, A GaAs Nanowire Array Solar Cell With 15.3% Efficiency at 1 Sun. *IEEE J. Photovolt.* **6**, 185–190 (2016).

48. Y. Su, *et al.*, Single-nanowire photoelectrochemistry. *Nat. Nanotechnol.* **11**, 609–612 (2016).

49. C. K. Chan, *et al.*, High-performance lithium battery anodes using silicon nanowires. *Nat. Nanotechnol.* **3**, 31–35 (2008).

50. C. K. Chan, R. N. Patel, M. J. O'Connell, B. A. Korgel, Y. Cui, Solution-Grown Silicon Nanowires for Lithium-Ion Battery Anodes. *ACS Nano* **4**, 1443–1450 (2010).

51. A. M. Chockla, *et al.*, Silicon Nanowire Fabric as a Lithium Ion Battery Electrode Material. *J. Am. Chem. Soc.* **133**, 20914–20921 (2011).

52. K. Hiruma, *et al.*, Growth and optical properties of nanometer-scale GaAs and InAs whiskers. *J. Appl. Phys.* **77**, 447–462 (1995).

53. H. Yu, J. Li, R. A. Loomis, L.-W. Wang, W. E. Buhro, Two- versus three-dimensional quantum confinement in indium phosphide wires and dots. *Nat. Mater.* **2**, 517–520 (2003).

54. T. Hanrath, B. A. Korgel, Nucleation and Growth of Germanium Nanowires Seeded by Organic Monolayer-Coated Gold Nanocrystals. *J. Am. Chem. Soc.* **124**, 1424–1429 (2002).

55. X. Chen, R. Liu, S. Qiao, J. Mao, X. Du, Synthesis of cadmium chalcogenides nanowires via laser-activated gold catalysts in solution. *Mater. Chem. Phys.* **212**, 408–414 (2018).

56. C. Huang, J. Mao, X. M. Chen, J. Yang, X. W. Du, Laser-activated gold catalysts for liquid-phase growth of cadmium selenide nanowires. *Chem. Commun.* **51**, 2145–2148 (2015).

57. X.-M. Chen, *et al.*, Laser-driven absorption/desorption of catalysts for producing nanowire arrays in solution. *J. Mater. Chem. A* **4**, 379–383 (2016).

58. J. Yeo, *et al.*, Laser-Induced Hydrothermal Growth of Heterogeneous Metal-Oxide Nanowire on Flexible Substrate by Laser Absorption Layer Design. *ACS Nano* **9**, 6059–6068 (2015).

59. M. J. Crane, *et al.*, Photothermal effects during nanodiamond synthesis from a carbon aerogel in a laser-heated diamond anvil cell. *Diam. Relat. Mater.* **87**, 134–142 (2018).

60. M. Kuno, An overview of solution-based semiconductor nanowires: synthesis and optical studies. *Phys. Chem. Chem. Phys.* **10**, 620–639 (2008).

61. T. Hanrath, B. A. Korgel, Supercritical Fluid–Liquid–Solid (SFLS) Synthesis of Si and Ge Nanowires Seeded by Colloidal Metal Nanocrystals. *Adv. Mater.* **15**, 437–440 (2003).

62. T. J. Trentler, *et al.*, Solution-Liquid-Solid Growth of Crystalline III-V Semiconductors: An Analogy to Vapor-Liquid-Solid Growth. *Science* **270**, 1791–1794 (1995).

63. H. Geaney, *et al.*, High Density Growth of Indium seeded Silicon Nanowires in the Vapor phase of a High Boiling Point Solvent. *Chem. Mater.* **24**, 2204–2210 (2012).

64. A. T. Heitsch, D. D. Fanfair, H.-Y. Tuan, B. A. Korgel, Solution−Liquid−Solid (SLS) Growth of Silicon Nanowires. *J. Am. Chem. Soc.* **130**, 5436–5437 (2008).

65. F. M. Davidson, R. Wiacek, B. A. Korgel, Supercritical Fluid−Liquid−Solid Synthesis of Gallium Phosphide Nanowires. *Chem. Mater.* **17**, 230–233 (2005).

66. A. M. Chockla, J. T. Harris, B. A. Korgel, Colloidal Synthesis of Germanium Nanorods. *Chem. Mater.* **23**, 1964–1970 (2011).

67. T. Hanrath, B. A. Korgel, Crystallography and Surface Faceting of Germanium Nanowires. *Small* **1**, 717–721 (2005).

68. X. Sun, G. Calebotta, B. Yu, G. Selvaduray, M. Meyyappan, Synthesis of germanium nanowires on insulator catalyzed by indium or antimony. *J. Vac. Sci. Technol. B Microelectron. Nanometer Struct. Process. Meas. Phenom.* **25**, 415–420 (2007).

69. M. J. Crane, X. Zhou, E. J. Davis, P. J. Pauzauskie, Photothermal Heating and Cooling of Nanostructures. *Chem. – Asian J.* **13**, 2575–2586 (2018).

70. E. J. Davis, Exact solutions for a class of heat and mass transfer problems. *Can. J. Chem. Eng.* **51**, 562–572 (1973).

71. S. W. Churchill, H. H. S. Chu, Correlating equations for laminar and turbulent free convection from a horizontal cylinder. *Int. J. Heat Mass Transf.* **18**, 1049–1053 (1975).

72. F. Wang, W. E. Buhro, An Easy Shortcut Synthesis of Size-Controlled Bismuth Nanoparticles and Their Use in the SLS Growth of High-Quality Colloidal Cadmium Selenide Quantum Wires. *Small* **6**, 573–581 (2010).

73. S. Wang, Y. Cheng, R. Wang, J. Sun, L. Gao, Highly Thermal Conductive Copper Nanowire Composites with Ultralow Loading: Toward Applications as Thermal Interface Materials. *ACS Appl. Mater. Interfaces* **6**, 6481–6486 (2014).

74. N. Balachander, *et al.*, Nanowire-filled polymer composites with ultrahigh thermal conductivity. *Appl. Phys. Lett.* **102**, 093117 (2013).

2828

# Laser-Driven Growth of Semiconductor Nanowires from Colloidal Nanocrystals via the Young-Laplace Effect

# SI Appendix

**Authors:** Elena P. Pandres[a,1], Matthew J. Crane[a,1], E. James Davis[a], Peter J. Pauzauskie[a,b,c,d,*], and Vincent C. Holmberg[a,c,e,*]

**Affiliations:**

[a]Department of Chemical Engineering, University of Washington, Seattle, WA 98195-1750, United States

[b]Department of Materials Science and Engineering, University of Washington, Seattle, WA 98195-2120, United States

[c]Molecular Engineering & Sciences Institute, University of Washington, Seattle, WA 98195-1652, United States

[d]Physical and Computational Sciences Directorate, Pacific Northwest National Laboratory, Richland, WA 99352, United States

[e]Clean Energy Institute, University of Washington, Seattle, WA 98195-1653, United States

[1]These authors contributed equally to this work

* Correspondence to holmvc@uw.edu and peterpz@uw.edu

Preparation of Cadmium Selenide Precursors

**Cadmium oleate**

Using standard Schlenk line techniques, 0.15 M cadmium oleate in 1-octadecene (ODE) was prepared by combining 210 mg CdO, 2.7 g oleic acid and 5.2 g ODE in a flask. The mixture was blanketed with nitrogen and stirred at 800 RPM, heated to 110°C, and placed under vacuum for 30 minutes. The mixture was cooled to room temperature under nitrogen and transferred to a glove box, whereupon 1 mL of trioctylphosphine (TOP) was added and mixed for 15 minutes.

**Trioctylphospine-selenide (TOP:Se)**

A 1.0 M stock solution of TOP:Se was prepared in a nitrogen-filled glove box by stirring 238 mg of Se in 5 mL of TOP overnight, until dissolved.

Nanowire Precursor Solutions for Photothermal Nanowire Growth

All nanowire precursor growth solutions were prepared and mixed under a nitrogen atmosphere in a quartz cuvette, which was lined with a thin layer of Teflon tape prior to sealing with a screw cap. The cap was then wrapped with Parafilm to avoid inadvertent nanocrystal oxidation.

**Bi-seeded cadmium selenide nanowire growth solution**

The Bi-seeded CdSe nanowire precursor growth solution was prepared under a nitrogen atmosphere in a quartz cuvette. 50 µL of a Bi nanocrystal dispersion (1.34 mg elemental Bi/L) in hexane, 125 µL of Cd-oleate (0.15 M in 1-ODE), and 213 µL of TOP:Se (1.0 M in 1-ODE) were combined and mixed in a screw-top quartz cuvette.

**Bi-seeded germanium nanowire growth solution**

The Bi-seeded Ge nanowire precursor growth solution was prepared under a nitrogen atmosphere in a quartz cuvette. 50 µL of a Bi nanocrystal dispersion (1.34 mg elemental Bi/L) in hexane, 100 µL of diphenylgermane (DPG), and 150 µL of squalane were combined and mixed in a screw-top quartz cuvette.

**In-seeded germanium nanowire growth solution**

The Bi-seeded Ge nanowire precursor growth solution was prepared under a nitrogen atmosphere in a quartz cuvette. 50 µL of an In nanocrystal dispersion in hexane, 100 µL of diphenylgermane (DPG), and 150 µL of squalane were combined and mixed in a screw-top quartz cuvette.

Continuous Flow Photothermal Nanowire Growth

For a typical continuous flow-based, photothermally driven nanowire growth synthesis, a glass tube with an inner diameter, outer diameter, and length of 4 mm, 5 mm, and 95 mm, respectively, was sealed under a nitrogen atmosphere with two small rubber septa. The reaction vessel was pre-filled with degassed 1-octadecene to displace all nitrogen prior to injection of the nanowire growth solution. The nanowire precursor growth solution, consisting of 60 µL of a Bi nanocrystal dispersion (1.34 mmol/L), 750 µL of Cd-oleate (0.15 M), 1.3 mL of TOP:Se (1.0 M), and 240 µL of degassed 1-ODE, was prepared in a nitrogen-filled glove box and loaded into a syringe. The needle of the precursor syringe was inserted through a rubber septum leading into the reaction vessel and the syringe was fixed onto a syringe pump in order to control the flow rate (5 mL/hr). A second needle was inserted into the other rubber septum on the exit side of the reaction vessel, leading to a glass vial used to collect the nanowire product. Prior to injecting the nanowire precursor growth solution, the laser was aligned to irradiate the center of the reaction vessel. After injecting the glass reactor vessel with the nanowire precursor solution, the reactor was irradiated with a polarized, near-infrared fiber laser ($\lambda$=1070 nm) at 5 W. The product was washed with a 2:1:1 ratio of toluene:chloroform:ethanol, collected by centrifugation at 19,000 RCF, dispersed into toluene, and washed two more times prior to sample characterization.

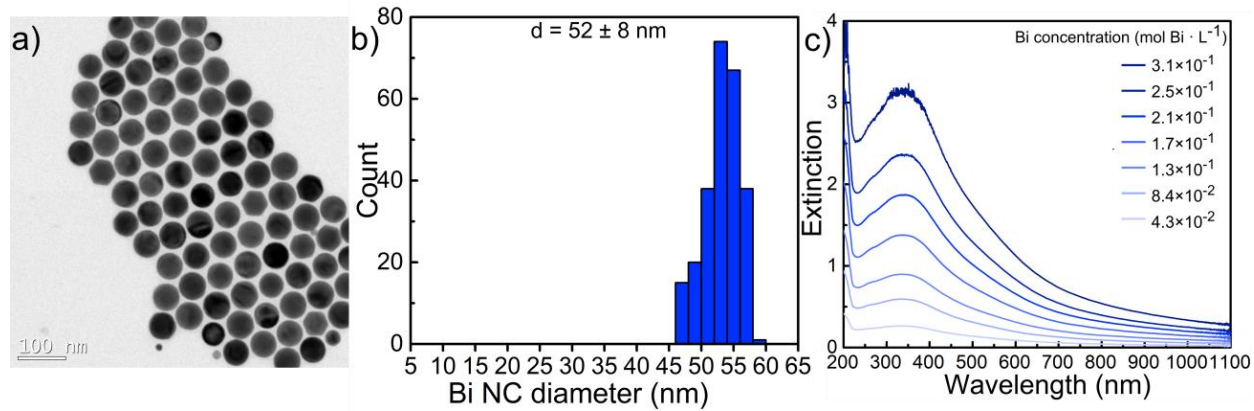

**Figure S1**. (a) TEM image of bismuth nanocrystals, (b) histogram of the bismuth nanocrystal diameter distribution, and (c) extinction spectra of the bismuth nanocrystals used to synthesize Bi-seeded Ge nanowires and CdSe nanowires.

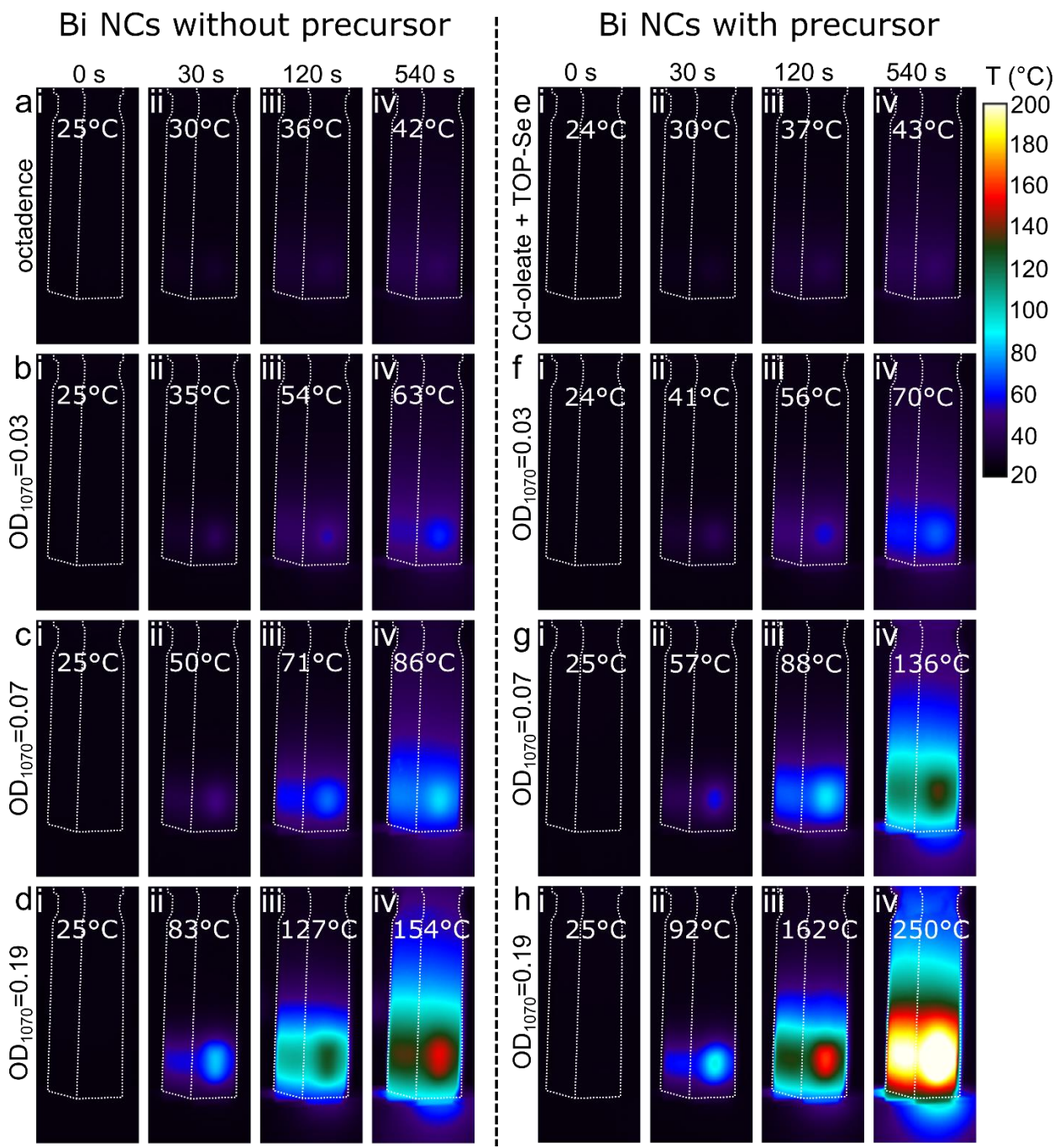

**Figure S2.** Thermographic infrared images of (a-d) different bismuth nanocrystal concentrations under irradiation (1070 nm, 15 W) in the absence of semiconductor precursors and (e-h) nanowire growth solutions with different nanocrystal concentrations under irradiation (1070 nm, 15 W), showing the evolution of the temperature profile in the cuvette.

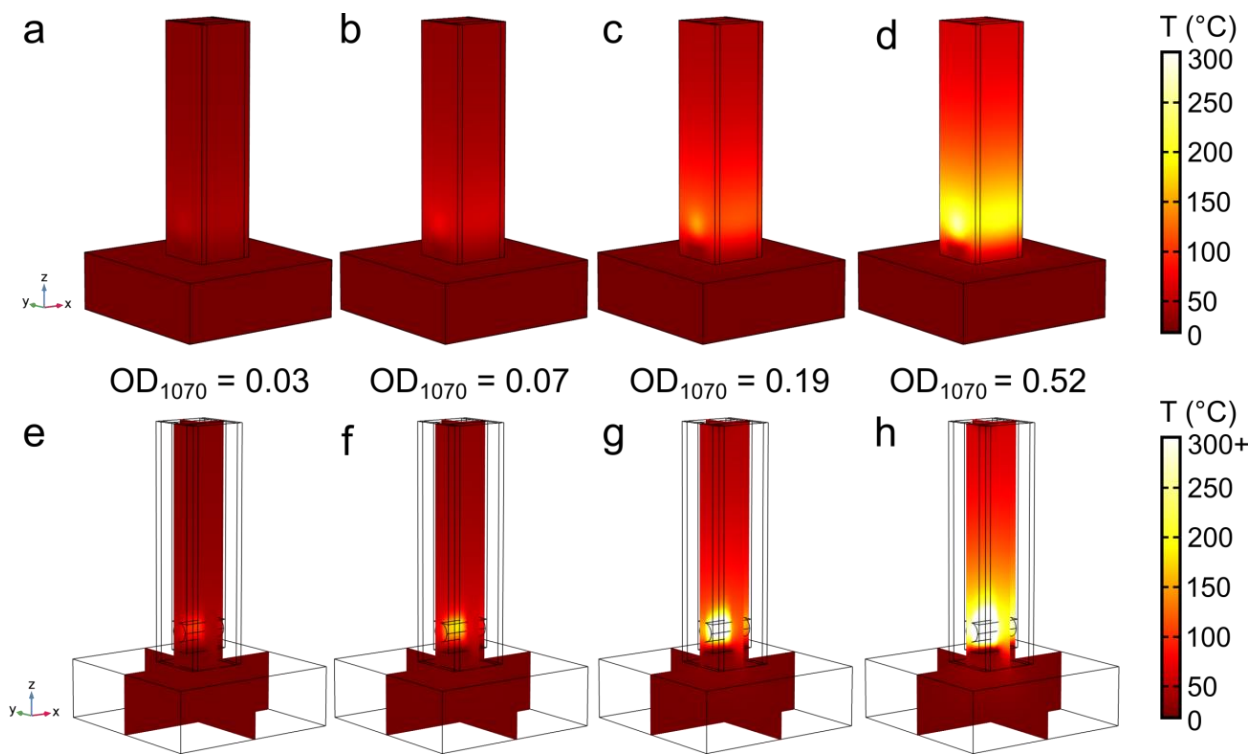

**Figure S3.** (a-d) Surface temperatures of quartz cuvettes with a range of Bi nanocrystal concentrations under 1070-nm irradiation, calculated via COMSOL modeling. (e-h) Cross-sectional temperature profiles of the above cuvettes calculated using COMSOL modeling.

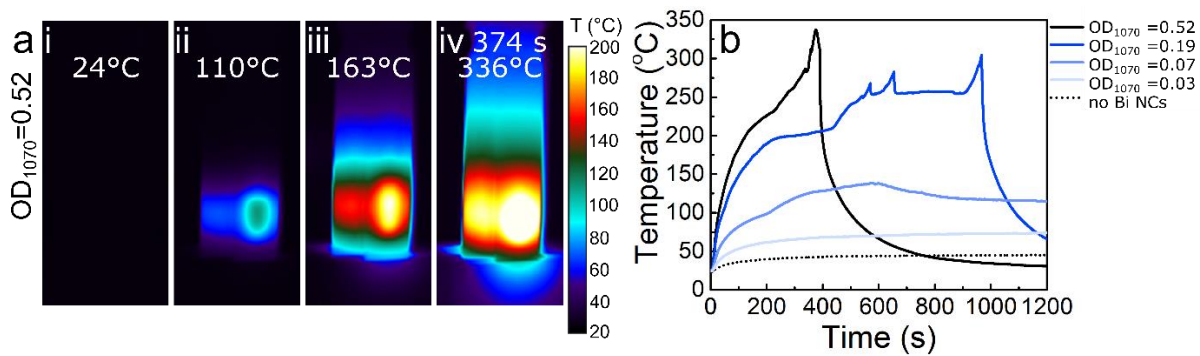

**Figure S4**. (a) Thermal infrared time series images of a nanowire precursor solution with bismuth nanocrystals at high concentrations ($OD_{1070}$=0.52) under irradiation (1070 nm, 15W), demonstrating the extremely high temperatures achieved by photothermal heating and (b) time-dependent temperature profiles for different bismuth nanocrystal concentrations ($OD_{1070}$=0.07-1.1) with CdSe precursors under irradiation (1070 nm, 15W).

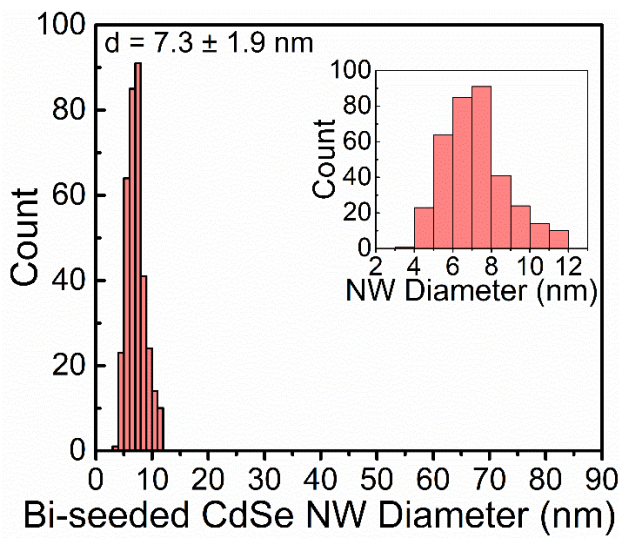

**Figure S5**. Diameter distribution of Bi-seeded CdSe nanowires synthesized by laser-driven growth.

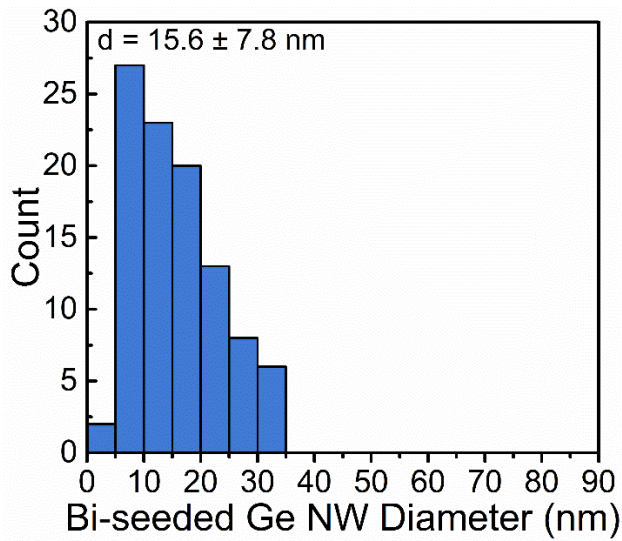

**Figure S6.** Diameter distribution of Bi-seeded germanium nanowires synthesized by laser-driven growth.

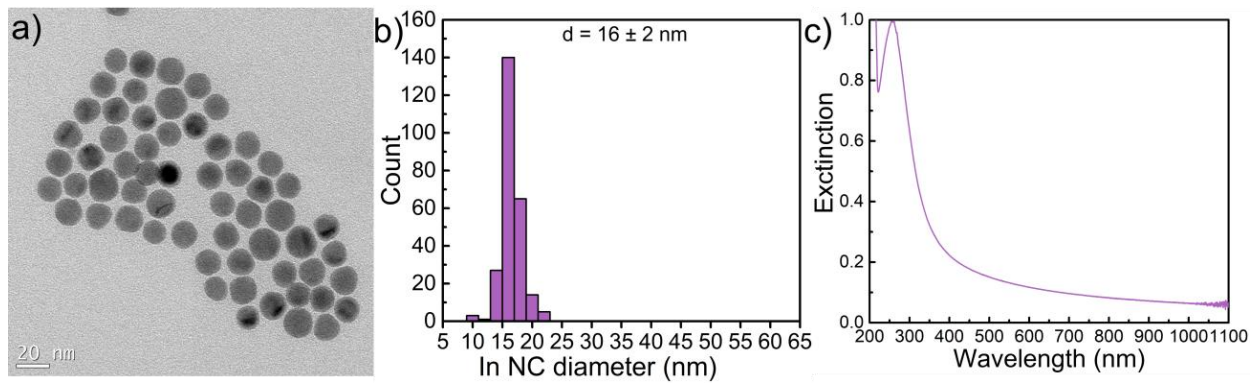

**Figure S7.** (a) TEM image of indium nanocrystals, (b) histogram of the indium nanocrystal diameter distribution, and (c) extinction spectrum of the indium nanocrystals used to synthesize In-seeded germanium nanowires.

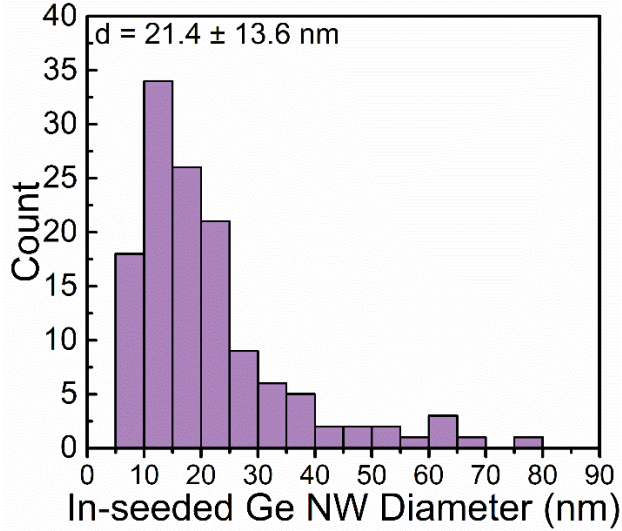

**Figure S8**. Diameter distribution of In-seeded germanium nanowires synthesized by laser-driven growth.

Analytical modeling of photothermally heated single nanocrystals and nanowires

To analyze heating of an isolated nanocrystal and metal-tipped nanowire under irradiation, we employed the analytical solutions outlined in previous publications.(1, 2) In the first step, we calculated the internal electric field amplitude in a spherical nanocrystal and in a metal-tipped nanowire, using Mie theory and the discrete dipole approximation implemented in Discrete Dipole Scattering (DDSCAT) software, respectively(3). We then converted these internal electric fields to volumetric source terms for heat transfer analysis. For the nanocrystal sphere, the heat transfer coefficient, $h$, is defined by the Nusselt number, $Nu = hL/k_s$, where $L$ is a characteristic length of the particle and $k_s$ is the solution thermal conductivity. For a sphere(1) suspended in an infinite bath, $Nu = 2.0$; for a vertical nanowire(4) suspended in an infinite bath, $Nu = 0.32$. In calculations of both nanocrystal (Fig. S9) and nanowire temperatures (Fig. S10), we used the highest irradiance and widest diameters. Because these length scales are far below the wavelength of light in any medium, there are no morphology-dependent resonances, and the predicted temperatures are maximized. Similarly, for the nanowire, we calculated the absorption for the largest projected nanowire surface area to achieve the largest volumetric source and averaged two orthogonal linear polarizations to simulate excitation by the near-infrared heating laser.

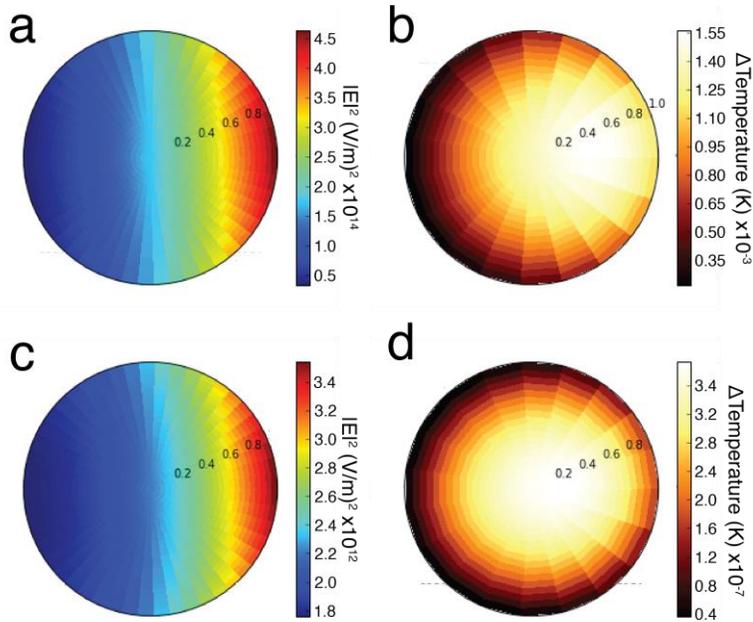

**Figure S9.** (a, c) Internal electric fields and (b, d) the resulting predicted temperature distributions within (a, b) a 50-nm diameter bismuth nanocrystal and (c, d) a 90-nm diameter indium nanocrystal. The irradiances for the bismuth and indium nanocrystal calculations were $3.2\times10^{-6}$ and $1.6\times10^{-6}$ W·m$^{-2}$, respectively.

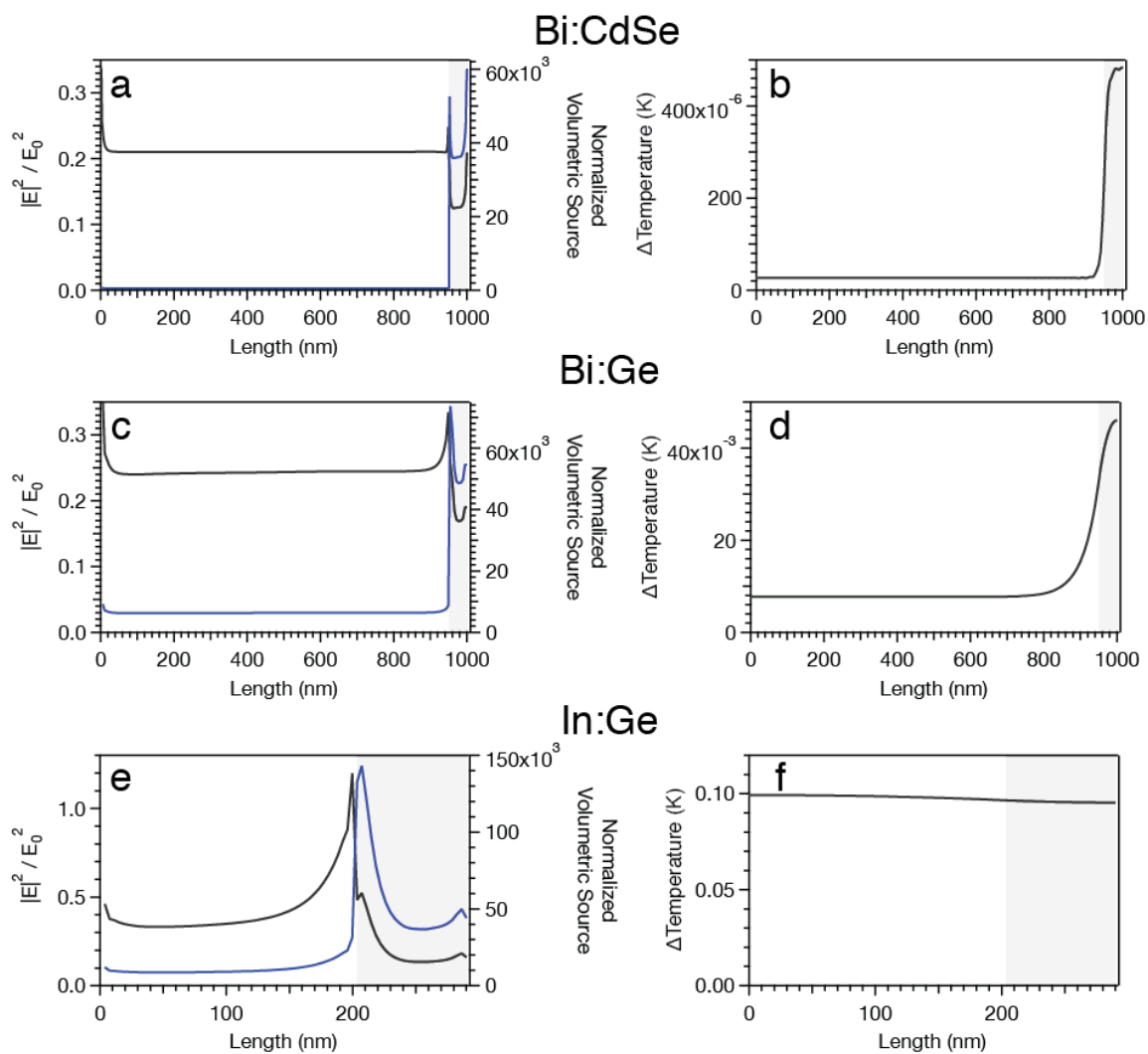

**Figure S10.** (a, c, e) Normalized internal electric field and volumetric source upon illumination of (a) a bismuth-seeded cadmium selenide nanowire with a 12-nm diameter and a 1000-nm length, (c) a bismuth-seeded germanium nanowire with a 40-nm diameter and a 1000-nm length, and (e) an indium-seeded germanium nanowire with an 80-nm diameter and a 290-nm length. The light grey areas illustrate the seed portion of each respective nanowire. (b, d, f) Calculated temperature changes along the length of the nanowire and seed for (b) bismuth-seeded cadmium selenide, (d) bismuth-seeded germanium, and (f) indium-seeded germanium nanowires.

Due to the larger complex refractive index of bismuth and indium compared to germanium and cadmium selenide, the magnitude of absorbed light and the resulting volumetric source terms are significantly larger in these regions. The increased magnitude of absorbed light correlates with higher increases in temperatures for (b) bismuth-seeded cadmium selenide and (d) bismuth-seeded germanium nanowires. Conversely, this effect is not observed in (e) indium-seeded germanium due to the high thermal conductivity of indium (f). The plotted volumetric sources were normalized to the volumetric source produced with E$_0$ = 1 V·m$^{-1}$. The irradiances for the bismuth and indium nanocrystal-seeded nanowire calculations were 3.2×10$^{-6}$ and 1.6×10$^{-6}$ W m$^{-2}$, respectively.

Heat transfer analysis of the cylindrical flow reactor

Fig. 4f-h shows infrared thermal images that demonstrate the unsteady-state development of the temperature distribution in the heated tube. To elucidate the effects of the numerous parameters on the heating process it is useful to analyze a quasi-steady-state model of the system. The energy equation that describes the temperature distribution is given by

$$2\rho_s C_s v_z \frac{\partial T}{\partial z} = k_s \left[ \frac{1}{r}\frac{\partial}{\partial r}\left(r \frac{\partial T}{\partial r}\right) + \frac{\partial^2 T}{\partial r^2} \right] + S, \qquad (1)$$

in which $\rho_s$, $C_s$, and $k_s$ are the density, specific heat, and the thermal conductivity of the fluid, respectively, $v_z$ is the velocity of the fluid, $r$ is the radial position, and $R_1$ is the inner radius of the tube ($R_1$ = 1.615 mm). The heat source function, S, is the rate of heat generation per unit volume due to the laser illumination over the region $z_1 \leq z \leq z_2$. The dominant resistance to heat transfer to the surroundings is that between the outer surface (at r = $R_2$ = 2 mm) and the surrounding air at temperature $T_\infty$, so the boundary condition at the surface is given by

$$-k_s \frac{\partial T}{\partial z}(R_2, z) = h[T(R_1, z) - T_\infty], \tag{2}$$

where the heat transfer coefficient is given by(4),

$$2\frac{R_2 h}{k_{air}} = 0.32. \tag{3}$$

There is a very extensive literature related to this type of problem since Graetz(5) analyzed heat transfer to a laminar flow in 1882. Consequently, problems of this type are called Graetz problems or extended Graetz problems depending on the boundary conditions and the inclusion of axial conduction. In the absence of axial conduction, analytical solutions of Graetz-like problems involving Kummer's function (the confluent hypergeometric function), have been developed by Davis(6). Papoutsakis and Ramkrishna(7) analyzed heat transfer in a capillary flow (Poiseuille flow) with the wall boundary condition given by Eq. (2), the Robin boundary condition. They solved the problem in the semi-infinite axial domain ($0 \le z \to \infty$) and pointed out the difficulties associated with this problem.

The velocity distribution at the inlet to the capillary tube is approximately uniform (plug flow), and far from the inlet Poiseuille flow occurs. We approximate the flow as plug flow here. It is convenient to write the energy equation in terms of dimensionless coordinates defined by

$$\xi = \frac{r}{R_1}, \zeta = \frac{(z-z_1)<v_z>}{R_1 Pe} \tag{4}$$

in which $<v_z>$ is the mean velocity, and the Peclet number is defined by

$$Pe = \frac{R_1 <v_z> \rho_s C_s}{k_s}. \tag{5}$$

Letting $\theta = T - T_\infty$ transforms the energy equation to give

$$\frac{\partial \theta}{\partial \zeta} = \frac{1}{\xi}\frac{\partial}{\partial \xi}\left(\xi \frac{\partial \theta}{\partial \xi}\right) + \frac{1}{Pe^2}\frac{\partial^2 \theta}{\partial \zeta^2} + \sigma, \tag{6}$$

in which the dimensionless heat source function is

$$\sigma = \frac{SR_1^2}{k_s}. \tag{7}$$

The wall boundary condition becomes

$$\frac{\partial \theta}{\partial \xi}(1,\zeta) = -Bi\,\theta(1,\zeta), \tag{8}$$

where the Biot number is

$$Bi = \frac{hR_1}{k_s} = \frac{0.16 k_{air} R_1}{R_2 k_s}. \tag{9}$$

For the geometry and properties of our capillary system we obtain $Pe = 1.81$. This low Peclet number indicates that there is significant axial conduction. Jerri and Davis(8) solved Eq. (6) for a constant wall temperature at $T > 0$, showing that there is a very large effect on the temperature distribution for $Pe < 10$. That large effect was also demonstrated by Papoutsakis and Ramkrishna(7) with Poiseuille flow for small Peclet numbers.

The heated region here extended from approximately $z_1 = 10.5$ mm to $z_2 = 14.5$ mm, or equivalently from $\zeta = 0$ to 1.77. Because the capillary tube was substantially longer (L = 40 mm) than the heated zone we consider three zones (i) $-\infty < \zeta \leq 0$ in which $S = 0$, (ii) $0 \leq \zeta \leq 1.77$, in which $S$ = constant, and (iii) $1.77 < \zeta < \infty$ in which $S = 0$.

At positions $\zeta = 0$ and $z_2$ we must apply compatibility conditions, that is, continuity of temperature and heat flux across these boundaries. These conditions are

$$\theta^-(\xi,0) = \theta^*(\xi,0), \quad \theta^*(\xi,0) = \theta^+(\xi,0),$$
$$\frac{\partial \theta^-}{\partial \zeta}(\xi,0) = \frac{\partial \theta^*}{\partial \zeta}(\xi,0), \quad \frac{\partial \theta^*}{\partial \zeta}(\xi,0) = \frac{\partial \theta^+}{\partial \zeta}(\xi,0). \tag{10}$$

In addition, we have the conditions

$$\theta^-(\xi,\zeta \to -\infty) = \text{bounded, and } \theta^+(\xi,\zeta \to -\infty) = \text{bounded}. \tag{11}$$

We look for solutions of the form

$$\theta(\xi,\zeta) = X(\xi)e^{(r\zeta)}, \tag{12}$$

where $X_j(\xi)$ are eigenfunctions defined by

$$X_j(\xi) = J_0(\lambda_j\xi) \text{ with } j = l, m, \text{ or } n, \tag{13}$$

in which the eigenvalues satisfy the transcendental equation

$$J_1(\lambda_j) - \frac{Bi}{\lambda_j}J_0(\lambda_j) = 0. \tag{14}$$

This yields the following solutions

$$\theta^- = \sum_{n=1}^{\infty} A_n^- e^{r_n\zeta} X_n(\lambda_n\xi) \tag{15}$$

with

$$r_n = \frac{Pe^2}{2}\left[1 - \sqrt{1 + 4\left(\lambda_n/Pe\right)^2}\right], \tag{16}$$

$$\theta^* = \sum_{m=1}^{\infty} [A_m^* e^{r_m\zeta} + B_m^* e^{r_n\zeta} + C_m]X_n(\lambda_n\xi), \tag{17}$$

and

$$\theta^+ = \sum_{n=1}^{\infty} A_l^+ e^{r_n\zeta} X_n(\lambda_n\xi) \tag{18}$$

in which

$$r_m = \frac{Pe^2}{2}\left[1 + \sqrt{1 + 4\left(\lambda_m/Pe\right)^2}\right], \tag{19}$$

and $I_m$ is the integral

$$I_m = \int_0^1 \xi \frac{X_m(\lambda_m\xi)}{\|X_m\|^2} d\xi, \tag{20}$$

where the norms squared are given by

$$\|X_m\|^2 = \int_0^1 \xi [X_m(\lambda_m \xi)]^2 d\xi \tag{21}$$

which is based on the orthogonality property

$$\int_0^1 \xi X_m(\lambda_m \xi) X_j(\lambda_j \xi) d\xi = \begin{cases} 0 \text{ for } j \neq m \\ \|X_m\| \text{ for } j = m \end{cases} \tag{22}$$

This orthogonality property can be used in the compatibility conditions to determine a set of equations relating the coefficients $A_n^-$, $A_m^*$, $B_m^*$, and $A_l^+$. For example, the first compatibility condition becomes

$$\theta^-(\xi, 0) = \sum_{n=1}^{\infty} A_n^- X_n(\lambda_n \xi) = \theta^*(\xi, 0) = \sum_{m=1}^{\infty} \left[ \left( A_m^* + B_m^* + \frac{\sigma I_m}{\lambda_m^2} \right) \right] X_m(\lambda_m \xi). \tag{23}$$

Applying the orthogonality of $X_n(\xi)$ to each term of the right-hand side of Eq. (23), one obtains

$$A_n^- \|X_n\|^2 = \left( A_m^* + B_m^* + \frac{\sigma I_m}{\lambda_m^2} \right) \int_0^1 \xi X_n(\xi) X_m(\xi) d\xi$$
$$= \left( \left( A_m^* + B_m^* + \frac{\sigma I_m}{\lambda_m^2} \right) \right) \|X_n\|^2, \tag{24}$$

with $n = m$.

The temperature distributions for the three zones are presented in Fig. 4 for various times and source functions based on the best-fit absorption efficiency, $\varepsilon_{abs}$. The source function $S$ in Eq. (1) is given by the product of the laser power, $P = 7.5$ W here, and the absorption efficiency of the system divided by the volume illuminated, that is,

$$S = \frac{P \, \varepsilon_{abs}}{\pi R_1^2 (z_2 - z_1)}. \tag{25}$$

The thermal conductivity of solutions containing dispersions of nanoparticles varies significantly with both concentration, morphology, and thermal conductivity of the constituent nanomaterials. For example, the thermal conductivity of polymer matrices have been demonstrated to increase more than 1000% upon adding copper nanowires at volume fractions equivalent to those in these experiments(9, 10). Thus, we varied the thermal conductivity of the nanowire solution from 0.10 to 1.0 W/mK to account for this potential variation. Table S1 lists the best-fit absorption efficiencies and thermal conductivities used to generate the calculated temperature profiles plotted in Fig. 4. Both the absorption efficiency and thermal conductivity were found to increase as the reaction proceeds, further confirming the photothermally driven nanowire reaction.

Applying orthogonality to solve for the coefficients $A_n^-$, $A_m^*$, $B_m^*$, and $A_l^+$ yields

$$A_n^- = \frac{C_m}{2}(1 - e^{-r_m \zeta_2})\left(\frac{r_n}{r_n - r_m}\right) \tag{26a}$$

$$A_m^* = -C_m e^{(-r_m \zeta_2)}\left(\frac{r_n}{r_n - r_m}\right) \tag{26b}$$

$$B_m^* = C_m \left(\frac{r_m}{r_n - r_m}\right) \tag{26c}$$

$$A_l^+ = C_m(1 - e^{(-r_n \zeta_2)})\left(\frac{r_m}{r_n - r_m}\right), \tag{26d}$$

where $C_m = \sigma I_m / \lambda_m^2$.

Table S1. Biot number, absorption efficiency, and thermal conductivity used for analytical heat transport fits in modeling the flow reactor.

| Heating Time (min) | Biot Number (-) | $\varepsilon_{abs}$ | $k_s$ |
|---|---|---|---|
| 1 | 0.042 | 0.248 | 0.100 |
| 2 | 0.034 | 0.327 | 0.123 |
| 5 | 0.005 | 0.494 | 0.860 |

# REFERENCES

1. M. J. Crane, *et al.*, Photothermal effects during nanodiamond synthesis from a carbon aerogel in a laser-heated diamond anvil cell. *Diam. Relat. Mater.* **87**, 134–142 (2018).

2. M. J. Crane, E. P. Pandres, E. J. Davis, V. C. Holmberg, P. J. Pauzauskie, Optically oriented attachment of nanoscale metal-semiconductor heterostructures in organic solvents via photonic nanosoldering. *Nat. Commun.* **10**, 1–7 (2019).

3. B. T. Draine, P. J. Flatau, Discrete-Dipole Approximation For Scattering Calculations. *J. Opt. Soc. Am. A* **11**, 1491 (1994).

4. P. B. Roder, B. E. Smith, E. J. Davis, P. J. Pauzauskie, Photothermal Heating of Nanowires. *J. Phys. Chem. C* **118**, 1407–1416 (2014).

5. L. Graetz, Ueber die Wärmeleitungsfähigkeit von Flüssigkeiten. *Ann. Phys.* **254**, 79–94 (1882).

6. E. J. Davis, Exact solutions for a class of heat and mass transfer problems. *Can. J. Chem. Eng.* **51**, 562–572 (1973).

7. E. Papoutsakis, D. Ramkrishna, Heat Transfer in a Capillary Flow Emerging from a Reservoir. *J. Heat Transf.* **103**, 429–435 (1981).

8. A. J. Jerri, E. J. Davis, Application of the sampling theorem to boundary value problems. *J. Eng. Math.* **8**, 1–8 (1974).

9. S. Wang, Y. Cheng, R. Wang, J. Sun, L. Gao, Highly Thermal Conductive Copper Nanowire Composites with Ultralow Loading: Toward Applications as Thermal Interface Materials. *ACS Appl. Mater. Interfaces* **6**, 6481–6486 (2014).

10. N. Balachander, *et al.*, Nanowire-filled polymer composites with ultrahigh thermal conductivity. *Appl. Phys. Lett.* **102**, 093117 (2013).